\documentclass[aps,pra,amsmath,amssymb,tightenlines,epsfig,floatfix,superscriptaddress, twocolumn, longbibliography, standalone]{revtex4-2}

\usepackage{soul}
\usepackage{graphicx}% Include figure files
\usepackage{subcaption}
\usepackage{tikz}
\usepackage{appendix}
\usepackage{physics}
\usepackage{float}
\usepackage{dcolumn}% Align table columns on decimal point
\usepackage{bm}
\usepackage{graphicx}% Include figure files
\usepackage{dcolumn}% Align table columns on decimal point
\usepackage{bm}% bold math
\usepackage{physics}
\usepackage{amsfonts}
\usepackage{xspace}
\usepackage{color}
\usepackage{xcolor}
\usepackage{epstopdf}
\usepackage{multirow}

\begin{document}
\newcommand{\psihat}{\ensuremath{\hat{\psi}}\xspace}
\newcommand{\psihatd}{\ensuremath{\hat{\psi}^{\dagger}}\xspace}
\newcommand{\ahat}{\ensuremath{\hat{a}}\xspace}
\newcommand{\Ham}{\ensuremath{\mathcal{H}}\xspace}
\newcommand{\ahatd}{\ensuremath{\hat{a}^{\dagger}}\xspace}
\newcommand{\bhat}{\ensuremath{\hat{b}}\xspace}
\newcommand{\bhatd}{\ensuremath{\hat{b}^{\dagger}}\xspace}
\newcommand{\boldr}{\ensuremath{\mathbf{r}}\xspace}
\newcommand{\dr}{\ensuremath{\,d^3\mathbf{r}}\xspace}
\newcommand{\dk}{\ensuremath{\,d^3\mathbf{k}}\xspace}
\newcommand{\etal}{\emph{et al.\/}\xspace}
\newcommand{\ie}{i.e.}
\newcommand{\eq}[1]{Eq.~(\ref{#1})\xspace}
\newcommand{\fig}[1]{Fig.~\ref{#1}\xspace}
\newcommand{\proj}[2]{\left| #1 \rangle\langle #2\right| \xspace}
\newcommand{\Qhat}{\ensuremath{\hat{Q}}\xspace}
\newcommand{\Qhatd}{\ensuremath{\hat{Q}^\dag}\xspace}
\newcommand{\phihatd}{\ensuremath{\hat{\phi}^{\dagger}}\xspace}
\newcommand{\phihat}{\ensuremath{\hat{\phi}}\xspace}
\newcommand{\boldk}{\ensuremath{\mathbf{k}}\xspace}
\newcommand{\boldp}{\ensuremath{\mathbf{p}}\xspace}
\newcommand{\boldsigma}{\ensuremath{\boldsymbol\sigma}\xspace}
\newcommand{\boldalpha}{\ensuremath{\boldsymbol\alpha}\xspace}
\newcommand{\parti}[2]{\frac{ \partial #1}{\partial #2} \xspace}
 \newcommand{\vs}[1]{\ensuremath{\boldsymbol{#1}}\xspace}
\renewcommand{\v}[1]{\ensuremath{\mathbf{#1}}\xspace}
\newcommand{\Psihat}{\ensuremath{\hat{\Psi}}\xspace}
\newcommand{\Psihatd}{\ensuremath{\hat{\Psi}^{\dagger}}\xspace}
\newcommand{\Vhatd}{\ensuremath{\hat{V}^{\dagger}}\xspace}
\newcommand{\Xhat}{\ensuremath{\hat{X}}\xspace}
\newcommand{\Xhatd}{\ensuremath{\hat{X}^{\dag}}\xspace}
\newcommand{\Yhat}{\ensuremath{\hat{Y}}\xspace}
\newcommand{\Jhat}{\ensuremath{\hat{J}}\xspace}
\newcommand{\Yhatd}{\ensuremath{\hat{Y}^{\dag}}\xspace}
\newcommand{\jhat}{\ensuremath{\hat{J}}\xspace}
\newcommand{\lhat}{\ensuremath{\hat{L}}\xspace}
\newcommand{\Nhat}{\ensuremath{\hat{N}}\xspace}
\newcommand{\rhohat}{\ensuremath{\hat{\rho}}\xspace}
\newcommand{\ddt}{\ensuremath{\frac{d}{dt}}\xspace}
\newcommand{\nset}{\ensuremath{n_1, n_2,\dots, n_k}\xspace}
\newcommand{\notes}[1]{{\color{blue}#1}}
\newcommand{\sah}[1]{\textcolor{purple}{#1}}
\newcommand{\jjh}[1]{\textcolor{teal}{#1}}
\newcommand{\tgmh}[1]{\textcolor{blue}{#1}}
\title{Calculating the quantum Fisher information via the truncated Wigner method}
% Force line breaks with \\

\author{Thakur G. M. Hiranandani}
\affiliation{School of Mathematica and Physics, University of Queensland, Brisbane, Australia}
\email{t.hiranandani@uq.edu.au}
\author{Joseph J. Hope} 
\affiliation{Department of Quantum Science and Technology, Research School of Physics, Australian National University, Canberra, Australia
}
\author{Simon A. Haine}
\affiliation{Department of Quantum Science and Technology, Research School of Physics, Australian National University, Canberra, Australia
}
\email{Simon.Haine@anu.edu.au}

\date{\today}

\begin{abstract}
In this work, we propose new methods of parameter estimation using stochastic sampling quantum phase-space simulations. We show that it is possible to compute the quantum Fisher information (QFI) from semiclassical stochastic samples using the Truncated Wigner Approximation (TWA). This method extends the class of quantum systems whose fundamental sensitivity limit can be computed efficiently to any system that can be modelled using the TWA, allowing the analysis of more meteorologically useful quantum states. We illustrate this approach with examples, including a system that evolves outside the spin-squeezing regime, where the method of moments fails.
\end{abstract}

\maketitle

\section{Introduction}
There is currently considerable interest in quantum sensing with entangled states, particularly atom interferometry \cite{Pezze_review:2018, Szigeti:2021, Greve:2022, Colombo:2022, Wilson:2024, Shen:2025, Sharma:2025, Cassens:2025}. In many cases, simple metrics such as the Wineland spin-squeezing parameter \cite{Wineland:1992} are sufficient to determine the metrological usefulness of the entanglement. However, the fundamental limit on the metrological potential of a given state is provided by the quantum Fisher information \cite{Braunstein:1994,Toth:2012, Toth:2014, Demkowicz-Dobrzanski:2014, Pezze_review:2018}. This more general quantity is useful in situations where the spin-squeezing parameter fails, either due to the non-Gaussian nature of the generated state \cite{Strobel:2014, Muessel:2015, Nolan:2017, Mirkhalaf:2018, Haine:2018b, Haine:2020, Haine:2021, Mehdi:2023}, or when a simple two-mode description is inadequate \cite{Haine:2016b, Kritsotakis:2018, Ben-Aicha:2024}.

The generation of many-particle entanglement is a result of inter-particle interactions, and therefore a theoretical model that can account for both the full quantum statistics, as well as accurately describe these interactions, is often required for a full understanding of the underlying physics \cite{Szigeti:2020}. Interacting many-body systems, such as dilute ultra-cold atomic gases or quantum-optical systems in nonlinear media, typically occupy Hilbert spaces that are too large to simulate directly, except in the cases where they can be modelled using a small number of modes. In regimes where spatial structure and quantum correlations are non-trivial, stochastic phase-space methods have  demonstrated significant success, and for atomic systems there has been widespread use of the truncated Wigner (TW) method in particular \cite{Drummond:1993, Steel:1998, Walls:2008, Sinatra:2002, Blakie:2008, Polkovnikov:2010, Ruostekoski:2013}. The TW method has been used to model the dynamics of Bose-Einstein condensates \cite{Steel:1998, Sinatra:1995, Norrie:2006, Drummond:2017}, including the generation of nonclassical correlations \cite{Haine:2014, Haine:2016b, Szigeti:2017, Szigeti:2020}, deleterious processes such as atomic scattering and phase-diffusion \cite{Haine:2011, Nolan:2016, Haine:2018}, atom-light entanglement \cite{Haine:2013, Szigeti:2014b, Haine:2016, Kritsotakis:2021, Fuderer:2023}, and feedback control \cite{Hush:2010, Zhu:2025}. These methods map the state of the system to a quasi-probability distribution and then calculate the evolution of that distribution via a stochastic sample. In phase-space methods, ensemble averages of stochastic samples provide estimates of the moments of the phase-space distribution \cite{Gardiner:2004b}. They are not efficient at reconstructing the full distribution, which has the same dimensionality as the original Hilbert space. This means that it is hard to calculate the fidelity between nearby states in regimes where we would otherwise wish to use the TW method. 

The quantum Fisher information is defined in terms of the infinitesimal distinguishability of nearby quantum states and therefore appears, at first sight, poorly suited to stochastic phase‑space approaches, since it depends on derivatives of the state with respect to the encoded parameter. Recently, RouhbakhshNabati \etal \cite{RouhbakhshNabati:2025} introduced a semiclassical phase‑space method for estimating the QFI based on the action accumulated along classical trajectories, and demonstrated its effectiveness in settings such as chaotic quantum systems where exact quantum simulations are intractable. While powerful in this context, that approach relies on the existence of a well‑defined classical action and is therefore not directly applicable in a number of experimentally relevant scenarios, including quantum fields (where the action is often zero, or independent of the parameter encoding dynamics) or protocols involving instantaneous parameter encoding. 

In this work, we present an alternative approach that is explicitly derived from the truncated Wigner approximation itself. We show that the QFI can be computed directly from the evolution of the sampled phase‑space trajectories and their parametric derivatives, without requiring reconstruction of the full quantum state or evaluation of classical actions. As a result, our method naturally integrates with existing TW simulations and is broadly applicable to interacting, spatially extended quantum systems relevant to contemporary quantum sensing experiments.

\section{Truncated Wigner Method and Metrological Information}
For a set of $k$ bosonic modes $\{\ahat_1, \ahat_2, \dots \ahat_k\}$, the density matrix is equivalent to a real-valued Wigner function of complex variables $\bm{\alpha}=[\alpha_1, \alpha_2, \dots \alpha_k]$:
\begin{subequations}
\begin{align}
W(\bm{\alpha}) &=\frac{1}{\pi^{k}} \mathrm{Tr}\left(\hat{w}^\dagger(\bm{\alpha}) \hat{\rho}\right)\,  
\end{align}
\begin{align}
\hat{\rho} &=\int d^{2k}\bm{\alpha}\; \hat{w}(\bm{\alpha})W(\bm{\alpha}) \, ,
\end{align}

\end{subequations}
where 
\begin{align}
\hat{w}(\bm{\alpha}) &= \frac{1}{\pi^{k}} \int d^{2k}\bm{\lambda}\;\exp(\bm{\lambda}\cdot\bm{\alpha}^*-\bm{\lambda}^*\cdot\bm{\alpha})\;\prod_{j=1}^k \hat{D}_j^\dagger (\lambda_j),
\end{align}
and $\hat{D}_j(\beta)=\exp(\beta \hat{a}_j^\dag - \beta^*\hat{a}_j)$
is the displacement operator for the $j$th mode \cite{Blakie:2008}. 

The equation of motion for the Wigner function $W(\bm{\alpha})$ for the system can be found from the master equation by using correspondences between differential operators on the Wigner function and the original quantum operators \cite{Walls:2008}. Specifically
\begin{subequations}
    \begin{align}
        \ahat_j\rhohat &\longleftrightarrow \left(\alpha_j + \frac{1}{2}\frac{\partial}{\partial \alpha_j^*}\right)W(\bm{\alpha}) \\
        \ahatd_j\rhohat &\longleftrightarrow \left(\alpha_j^* - \frac{1}{2}\frac{\partial}{\partial \alpha_j}\right)W(\bm{\alpha}) \\
        \rhohat\ahat_j &\longleftrightarrow \left(\alpha_j - \frac{1}{2}\frac{\partial}{\partial \alpha_j^*}\right)W(\bm{\alpha}) \\
        \rhohat\ahatd_j &\longleftrightarrow \left(\alpha_j^* + \frac{1}{2}\frac{\partial}{\partial \alpha_j}\right)W(\bm{\alpha})
    \end{align}
    \label{wig_correspondences}
\end{subequations}
The Truncated Wigner (TW) approximation is where we truncate the third- and higher-order derivatives of the Wigner function's equation of motion, which results in a Fokker-Planck equation (FPE). The FPE can then be mapped to a set of stochastic differential equations for complex variables $\bm{\alpha}$, which are initialised by a statistical sample of the initial Wigner function. A single stochastic solution of these equations is called a trajectory. This method is computationally easier than working with the full Wigner function, or equivalently, the full quantum state, as evolving stochastically sampled trajectories requires logarithmically less memory. 
 
The TW method can be used to calculate moments of the probability distribution. Specifically, expectation values and other moments are calculated via the mapping 
\begin{align}
\langle :f(\mathbf{\ahat}, \mathbf{\ahat}^\dag):_\mathrm{sym}\rangle &= \int d^{2k}\bm{\alpha}\; f(\bm{\alpha},\bm{\alpha}^*) W(\bm{\alpha},t) \nonumber  \\ 
&= \mathbb{E}\left[f(\bm{\alpha}(t),\bm{\alpha}^*(t))\right], \label{TWcorrespondences}
\end{align}
where `sym' denotes symmetric ordering and $\mathbb{E}\left[\dots\right]$ denotes the mean over many stochastic trajectories.

For many applications in quantum sensing it is desirable to calculate the quantum Fisher information, which provides a limit on the precision of parameter estimation through the quantum Cramer-Rao bound: $\Delta \omega \geq F_Q^{-\frac{1}{2}}$, where $\omega$ is the metrological parameter of interest \cite{Braunstein:1994, Pezze_review:2018}. When the parameter is encoded onto some quantum state $\hat{\rho}$, this can be calculated via
\begin{align}
    F_Q &= \Tr \left[\hat{\rho}(\omega) \hat{L}^2_\omega \right]
\end{align}
where the $\hat{L}$ is the symmetric logarithmic derivative defined implicitly as 
\begin{align}
    \partial_\omega \hat{\rho} = \frac{1}{2}\left(\hat{\rho} \hat{L}_\omega + \hat{L}_\omega\hat{\rho} \right).
\end{align}
For pure states such that $\hat{\rho}^2 = \hat{\rho}$, this simplifies to
 \begin{align}
    F_Q &= 2 \Tr\left[\left(\partial_\omega  \hat{\rho} \right)^2\right]. \label{FQpure}
\end{align}

Assuming the system remains pure, the most general form of parameter encoding is via some unitary dynamics such that
\begin{align}
\rhohat = \hat{U}_\omega \rhohat_0 \hat{U}_\omega^\dag \, ,
\end{align}
where $\hat{U}_\omega$ is a general unitary operator that depends on the parameter $\omega$. In some cases $\hat{U}_\omega$ can be expressed in the simple form
\begin{align}
\hat{U} = \exp\left(-i\omega \hat{G}\right) \label{unitary_encoding}
\end{align}
for some known Hermitian operator $\hat{G}$, in which case 
\begin{align}
F_Q &= 4 \mathrm{Var}(\hat{G})
\end{align}
where the expectation value is taken with respect to $\rhohat_0$. In this case, for operators $\hat{G}$ that can be expressed in terms of creation and annihilation operators, it is straightforward to calculate the QFI using the TW method, by simply evaluating $\langle \hat{G}\rangle$ and $\langle \hat{G}^2\rangle$ using stochastic averages.

However, in general, not all encoding can be easily expressed in the form of \eq{unitary_encoding}. A common example is where the Hamiltonian is of the form $\hat{A}(t) + \omega \hat{B}(t)$ where $\hat{A}$ and $\hat{B}$ do not commute. In this case, once the parameter is encoded in the state, we must use \eq{FQpure} to express the QFI. We can write this in terms of the Wigner function instead of the density operator, which allows us to express it in terms of stochastic averages of trajectories:
\begin{subequations}
\begin{align}
F_Q &= 2 \;\pi^k\int d^{2k}\alpha \left(\partial_\omega W(\bm{\alpha},\omega,t)\right)^2  \label{FQ_Wig_full} \\
&=  2 \;\pi^k \;\mathbb{E}\left[\left(\frac{\left(\partial_\omega W(\bm{\alpha}(\omega,t),\omega,t)\right)^2}{W(\bm{\alpha}(\omega,t),\omega,t)}\right)\right]. \label{FQ_wig_int}
\end{align}
\end{subequations}

Even though Eq.~\ref{FQ_wig_int} is in terms of a stochastic average of the trajectories $\bm{\alpha}(\omega,t)$ which can be efficient to compute, it also requires knowledge of the full Wigner function and its derivative. Unfortunately, we are mainly interested in situations where the Wigner function is not known analytically. 

The naive approach, where the Wigner function is computed by sampling trajectories for two nearby values of $\omega$, is computationally equivalent to direct simulation in the full Hilbert space. Thus, the stochastic unravelling is unsurprisingly very susceptible to sampling error.
We have found that even for single-mode systems, this approach requires orders of magnitude more trajectories than is required to reliably estimate expectation values and other low-order moments. Fortunately we can exploit \eq{FQ_wig_int} to construct a more direct calculation of the derivative using the stochastic unravelling. 

The exact Wigner equation follows the Moyal equation, which follows from noting that the Wigner function is effectively the Wigner-Weyl transformation of the density matrix \cite{Schroeck:1996}. For Hamiltonian evolution, the TW approximation is equivalent to expanding the Moyal bracket to the lowest order of derivatives, which is roughly equivalent to the lowest order in $\hbar/S$, where $S$ is the classical action \cite{Polkovnikov:2010}. In this case the evolution of the Wigner function is defined by symplectic evolution \cite{Polkovnikov:2010}:
\begin{align}
    \label{eq:TW_equation}
    i \frac{\partial W}{\partial t} = \frac{\partial W_H}{\partial \bm{\alpha}} \cdot  \frac{\partial W}{\partial \bm{\alpha}^*} - \frac{\partial W_H}{\partial \bm{\alpha}^*}\cdot \frac{\partial W}{\partial \bm{\alpha}}
\end{align}
where $W_H$ is the Wigner transform of the Hamiltonian $\hat{H}(\omega)$. Eq.~\eqref{eq:TW_equation} has the solution 
\begin{align}
    W(\bm{\alpha},\omega,t) = W_0(\bm{\alpha}(\omega,-t)),
\end{align}
where the trajectories $\bm{\alpha}=\bm{\alpha}(\omega,0)$ sample the initial Wigner function $W_0$, and their time-reversed evolution,
\begin{align}
    \frac{d}{dt}\bm{\alpha}(\omega, t) = - i \frac{\partial W_H}{\partial \bm{\alpha}^*}
\end{align}
samples the evolved Wigner function.
This immediately equates the denominator of \eq{FQ_wig_int} with $W_0(\bm{\alpha}(\omega,0))$, as the evolution of the trajectories effectively cancel the evolution of the Wigner function. The numerator of the QFI contains the derivative of the Wigner function with respect to $\omega$, which we can write in terms of the derivatives of the trajectories and the initial Wigner function:
\begin{align}
\partial_\omega W(\mathbf{x},\omega,t) = \partial_\omega \mathbf{x}(\omega,-t) \cdot \nabla_\mathbf{x} W_0(\mathbf{x}(\omega,-t)) \, \label{dwW},
\end{align}
where we write the set of $k$ complex variables $\bm{\alpha}$ as a set of $2k$ real variables $\mathbf{x}$, as the Wigner function is not complex analytic. Typically, these variables would be real and imaginary components of $\alpha_j$, or the quadratures $X_j = \frac{1}{\sqrt{2}}(\alpha_j + \alpha_j^*)$, $Y_j = \frac{i}{\sqrt{2}}(\alpha_j - \alpha_j^*)$.
We require the value of this derivative evaluated at the evolved trajectories $\partial_\omega W(\bm{\alpha}(\omega,t),\omega,t)$, so again we find that the evolution of the trajectories cancels the time dependence of the function, and we can calculate \eq{dwW} using:
\begin{align}
\partial_\omega W(\mathbf{x}(\omega,t),\omega,t) = \partial_\omega \mathbf{x}(\omega,0) \cdot \nabla_\mathbf{x} W_0(\mathbf{x}(\omega,0)) \, \label{dwWasReq}.
\end{align}
The gradient of the initial Wigner function can be computed analytically and evaluated at $\mathbf{x}(\omega,0)$, which is simply the initial sample point for the trajectory. The function $\partial_\omega \mathbf{x}(\omega,0)$ can be calculated by evolving the trajectory forward in time using two values for $\omega$ near the operating point, mapping both back using the default value and then using a finite difference to estimate the derivative. Rather than attempting to reconstruct the Wigner function and then model its dependence on $\omega$, this method samples that derivative directly through the $\omega$-dependence of each trajectory. 

While the method described above is general, in practice, the truncated Wigner method almost always utilises Gaussian initial states \cite{Olsen:2009}. Expressing $W_0(\mathbf{x})$ in the most general Gaussian form
\begin{align}
    \label{eq:GaussianWignerFunction}
    W_0(\mathbf{x}) = \frac{1}{\pi^{k}} \exp\left(-(\mathbf{x}-\bm{\mu})^T \mathbf{M}^{-1} (\mathbf{x}-\bm{\mu}) \right),
\end{align}
where $\bm{\mu}$ and $\mathbf{M}$ are the vector of means and covariance matrix of $\mathbf{x}$. For the vacuum and coherent states $\mathbf{M} = \mathbf{I}_{2k}$. The gradient is hence given by
\begin{align}
    \frac{\partial W_0(\mathbf{x})}{\partial \mathbf{x}} &= - \mathbf{M}^{-1} (\mathbf{x} - \bm{\mu}) W_0(\mathbf{x}),
\end{align}
and \eq{FQ_wig_int} becomes
\begin{align}
    F_Q &= 2 \pi^k\mathbb{E}\left[ W_0(\mathbf{x})\left(
    \partial_\omega \mathbf{x}\cdot \mathbf{M}^{-1}(\mathbf{x} -\bm{\mu})\right)^2\right].
    \label{eq:TWA_QFI_EXP}
\end{align}

\section{Illustrative examples:}
We now demonstrate our method using a simple single-mode example. We begin with the well-studied model of parametric amplification of an optical cavity mode, in a perfect (lossless) optical cavity. We first consider the system in the undepleted pump approximation, where there exists an analytic solution, and then demonstrate the utility of our scheme by introducing depletion from the pump mode. 
\subsection{Undepleted pump approximation}
We aim to calculate the QFI of the state resulting from evolution under the Hamiltonian: 
\begin{align}
\hat{H}_\mathrm{OPO} &= \frac{\hbar g}{2} \left(\ahatd\ahatd e^{i\theta} + \ahat\ahat e^{-i\theta} \right) \,
\end{align}
for duration $t_1$, before the parameter $\omega$ is encoded via evolution under the Hamiltonian: 
\begin{align}
\hat{H}_\mathrm{\omega} = \hbar \omega \ahatd\ahat \label{Homega}
\end{align}
for duration $\Delta t = t_2-t_1$. Physically, $\hat{H}_\mathrm{OPO}$ represents the Hamiltonian for optical parametric amplification in the undepleted pump approximation.  
Using \eq{wig_correspondences}, the von-Neumann equation $i\hbar \ddt \rhohat = \left[\hat{H}_\mathrm{OPO},\rhohat\right]$ maps to
\begin{align}
\ddt W(\alpha) &= ig\left(e^{i\theta} \frac{\partial}{\partial \alpha}\left(\alpha^* W(\alpha)\right) - \mathrm{c.c.}\right)
\end{align}
which is of the form of an FPE, so maps exactly to the equivalent ordinary differential equation (ODE):
\begin{align}
i \ddt \alpha &= g e^{i \theta}\alpha^* \, ,
\end{align} 
which has solution
\begin{align}
\alpha(t_1) &= \cosh(gt_1) \alpha(0) -i e^{i\theta} \sinh(gt_1) \alpha^*(0) \, \label{eaxmple1prep}.
\end{align}
Similarly, evolution under $\hat{H}_\mathrm{OPO}$ results in the equation of motion
\begin{align}
i \ddt \alpha &= \omega \alpha
\end{align}
and is solved by
\begin{align}
\alpha(t_2) &= \alpha(t_1) e^{-i\omega \Delta t} \, .
\end{align}
We choose the initial state to be a coherent state $|\alpha_0\rangle$, so initial conditions for each trajectory are stochastically sampled from the Wigner distribution for this state, such that $\alpha(0) = \alpha_0 + \eta$, where $\eta$ is complex Gaussian noise with zero mean and $\mathbb{E}(\abs{\eta}^2) = \frac{1}{2}$. 

In order to calculate the QFI at the operating point $\omega=0$ and time $t_2$, we first integrate the same initial conditions using two nearby values: $\omega=\pm \frac{1}{2}\Delta\omega$. We then rewind both trajectories using $\omega=0$. This allows us to use finite difference to calculate the derivative of the real and imaginary components of $\alpha$ at $t=0$, as required in equation \ref{dwWasReq}. In this example, all these quantities can be computed analytically, though in general this process will require numerical integration. We then repeat this over many trajectories, and compute the QFI from the stochastic average as given in \eq{eq:TWA_QFI_EXP}.

\begin{figure}[h]
\begin{center}
\includegraphics[width=\columnwidth]{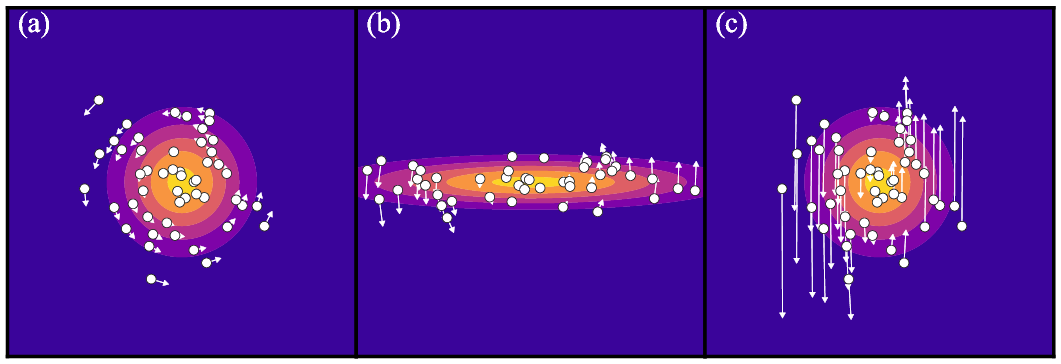}
\caption{A small subsample of individual trajectories (white dots) overlaid with the analytic form of the Wigner function for (a) an initial vacuum state, (b) the state at time $t_1$, and (c) the time-reversed state after encoding. The arrows indicate the derivative of the trajectories w.r.t. to $\omega$, or `flow'. The state preparation breaks the rotational symmetry of the state, so the state is changed by the encoding, so the QFI becomes non-zero. When the trajectories from (b) are mapped to $t=0$, we can compute this non-zero QFI directly from the trajectories, as the flow has gained a component in the direction of the initial gradient.}
\label{trajectories}
\end{center}
\end{figure}

Figure \ref{trajectories} illustrates how the preparation phase affects the QFI for an initial vacuum, both in terms of the Wigner function and our trajectory method. The derivative of each trajectory with respect to $\omega$ is indicated with arrows. In (a) we see that the initial state has precisely zero QFI as the encoding does not change the vacuum. The individual trajectories rotate due to $\omega$, but this flow is exactly parallel to the contours of $W$, (or equivalently, perpendicular to the gradient of $W$), so \eq{dwWasReq} shows they all give zero contribution to the QFI. At time $t_2$, after the application of $\hat{H}_\mathrm{OPO}$, the Wigner function has changed shape, and the flow of trajectories shows that its orientation is clearly dependent on $\omega$. When these trajectories are reversed in time to $t=0$, the flow field has changed shape such that they are no longer perpendicular to the gradient, giving a non-zero QFI. 

\begin{figure}[h]
%\begin{center}
\includegraphics[width=\columnwidth]{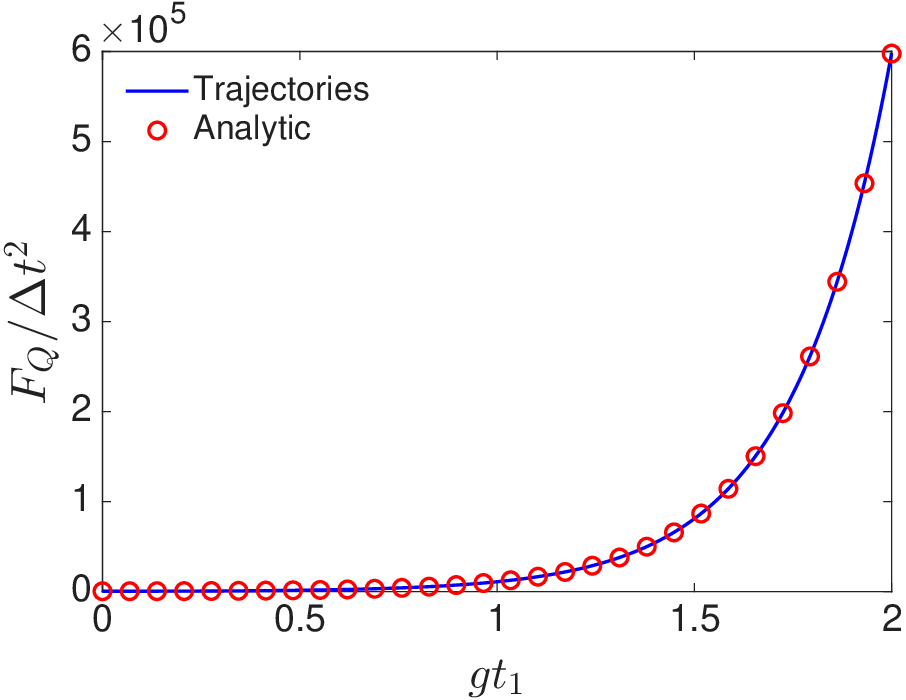}
\caption{QFI calculated from \eq{eq:TWA_QFI_EXP} (blue line) compared to the analytic solution \eq{FQ_opo_anal} (red circles). 1 million trajectories were used. Parameters: $\alpha_0=10$, $\theta=0$.}
\label{Example1Agreement}
%\end{center}
\end{figure}

Figure \ref{Example1Agreement} shows the QFI calculated from \eq{FQ_wig_int} for a non-zero value of $\alpha_0$. We see perfect agreement with the analytic solution for all values of $gt$, $\alpha_0$, and $\theta$:  
\begin{align}
\frac{F_Q}{\Delta t^2} &= 4 \mathrm{Var}(\ahatd(t_1)\ahat(t_1)) \nonumber \\
&= \left(4|\alpha_0|^2 +1\right)\cosh 4 gt_1 \nonumber \\
&- 4|\alpha_0|^2\sin\left(2\vartheta -\theta\right)\sinh 4gt_1 -1 \label{FQ_opo_anal}
\end{align}
for $\alpha_0 = |\alpha_0| e^{i\vartheta}$. 

\begin{figure}[h]
\begin{center}
\includegraphics[width=\columnwidth]{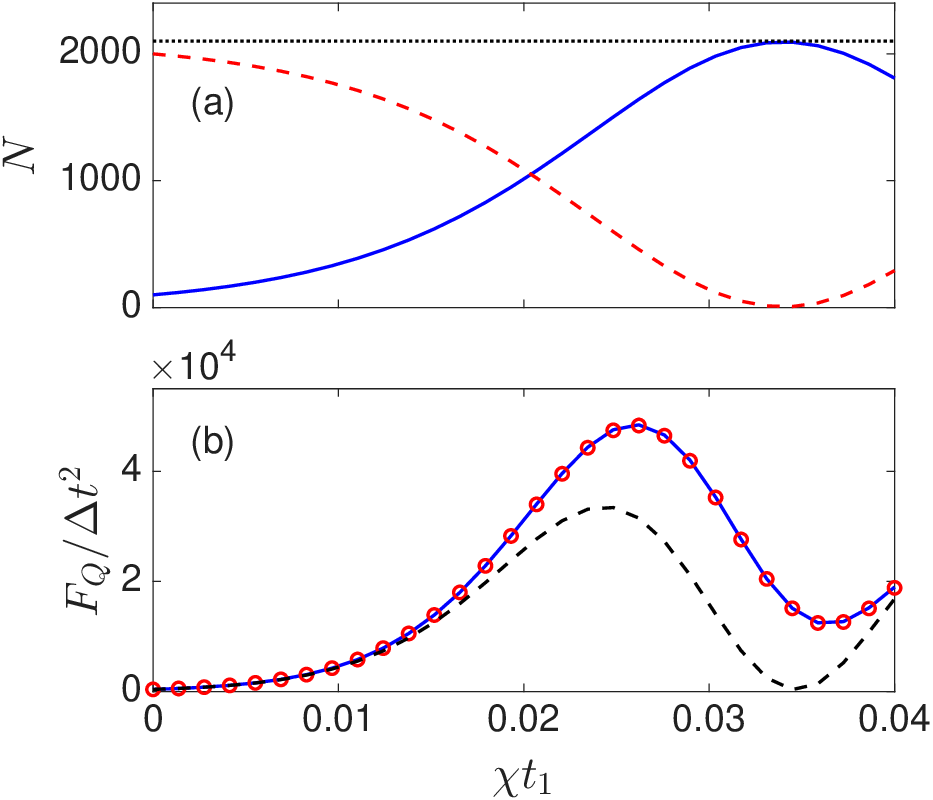}
\caption{(a) Populations of cavity and pump modes during state preparation, following \eq{undepeletedTrajectories}. Blue line: $N_a = \langle \ahatd(t_1)\ahat(t_1)\rangle$, red dashed line $2N_b = 2\langle\bhatd(t_1)\bhat(t_1)\rangle$, black dotted line: $N_a + 2N_b$. (b) QFI calculated from \eq{eq:TWA_QFI_EXP} (blue line), QFI calculated from variance $4 \mathrm{Var}(\ahatd(t_1)\ahat(t_1))$ (red circles), QFI contribution from the $\partial_\omega \alpha$ terms (black dashed line). All three traces were evaluated via the TW method, using 1 million trajectories. Parameters: $\alpha_0=10$, $\beta_0 = \sqrt{1000}$, $\theta=0$.}
\label{fig:Hchi}
\end{center}
\end{figure}

\subsection{Effects of pump depletion}
As a more involved example, we include the effects of depletion from the pump mode. In this case, the dynamics cannot be solved analytically, and a numeric method, such as TW is required \cite{Haine:2013}. Introducing $\hat{b}$ as the annihilation operator for the pump mode, we consider the Hamiltonian
\begin{align}
\hat{H}_\chi &= \frac{\hbar \chi}{2}\left(\ahatd\ahatd \bhat e^{i\theta} + e^{-i\theta}\bhatd\ahat\ahat\right) \, . \label{Hchi}
\end{align}
Using the operator correspondences, this maps to the equation of motion for the Wigner function
\begin{align}
\ddt W(\alpha, \beta) &= i\chi e^{i\theta} \left( \frac{\partial}{\partial \alpha}\left(\alpha^*\beta W(\alpha,\beta)\right) \right. \notag \\
&+ \frac{1}{2}\frac{\partial}{\partial \beta}\left(\alpha^*\alpha^* W(\alpha,\beta)\right)  \notag \\
&+ \left. \frac{1}{8}\frac{\partial}{\partial \beta}\frac{\partial}{\partial \alpha^*}\frac{\partial}{\partial \alpha^*} W(\alpha, \beta) \right) + \mathrm{c.c}. 
\end{align}
Unlike the previous example, this equation is not an exact FPE, due to the existence of the third-order derivatives. In order to map this to an ordinary differential equation (ODE), we ignore these terms, with the justification that for short times, their effect is significantly less than the first-order terms. Additionally, for large mode occupation, as is often considered in interacting systems of ultra-cold atoms, such as Bose-Einstein condensates, the first-order terms are larger by approximately a factor of the mode-occupation. Neglecting these terms, we obtain
\begin{subequations}
\begin{align}
i\ddt \alpha &= \chi e^{i\theta} \beta \alpha^* \\
i\ddt \beta &= \frac{\chi}{2}e^{-i\theta} \alpha^2 \label{undepeletedTrajectories}
\end{align}
\end{subequations}
These equations do not have an analytic solution, so are solved numerically. As before, after evolving under $\hat{H}_\chi$ followed by the parameter encoding under $\hat{H}_\omega$, we reverse the dynamics of $\hat{H}_\chi$ in order to calculate the QFI. Figure \ref{fig:Hchi} shows the QFI calculated from this method, compared to $F_Q = 4\Delta t^2 \mathrm{Var}(\ahatd(t_1)\ahat(t_1))$, also calculated via the TW method. 

We see perfect agreement between the two methods of computing the QFI. However, both of these have made the same approximation, that is, neglecting the 3rd-order derivative terms in the Wigner function dynamics. We also show the QFI calculated using only the $\partial_\omega \alpha$ terms. While the phase shift is applied solely to the mode represented by $\alpha$, in the limit of large depletion there is considerable entanglement with the pump mode. This means that the $\partial_\omega \beta$ terms contain a significant fraction of the total QFI. 

\subsection{Kerr-Interaction}
\begin{figure}[h]
\begin{center}
\includegraphics[width=\columnwidth]{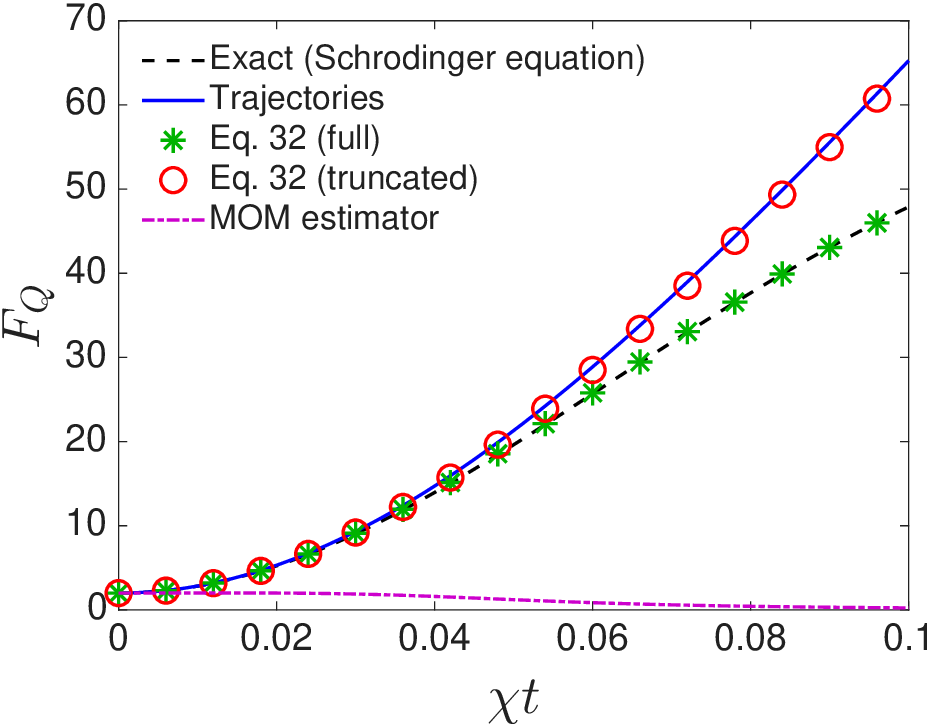}
\caption{$F_Q$ calculated from the trajectory method (blue soild line), compared to the exact solution from the Schrodinger equation (black dashed line), and from the solution to \eq{FP_Kerr}, both with (green stars) and without (red circles) the inclusion of the third-order derivative terms. The equivalent sensitivity metric from the MoM estimator $1/(\delta x_0)^2 = 1/\mathrm{Var}(\hat{X})$, calculated via exaction solution from the Schrodinger equation, is shown with the orange dot-dashed line. An initial coherent state $|\alpha_0\rangle$ with $\alpha_0 = 4$ was used. }
\label{fig:kerrFQ}
\end{center}
\end{figure}

We now consider the dynamics generated by the well-known Kerr Hamiltonian:
\begin{align}
\hat{H}_\mathrm{Kerr} &= \frac{\hbar \chi}{2} \ahatd\ahatd\ahat\ahat - \hbar \omega_0 \ahatd\ahat \, , \label{Hkerr}
\end{align}
where the term proportional to $\omega_0$ is included for simplicity to remove the bulk rotation of the state in the $X-Y$ plane. This Hamiltonian represents the dynamics that induce squeezing for propagation in a nonlinear refractive index \cite{Bachor:2004}. Using the operator correspondences, \eq{Hkerr} maps to
\begin{align}
\frac{d}{dt} W(\alpha) &= i\frac{\partial}{\partial \alpha}\left[\chi\left(|\alpha |^2 -1\right) -\omega_0\right]\alpha W(\alpha) \notag \\
&+ \frac{i}{4}\frac{\partial}{\partial \alpha}\frac{\partial}{\partial \alpha}\frac{\partial}{\partial \alpha^*}\alpha W(\alpha) + \mathrm{c.c.} \label{FP_Kerr}
\end{align}
In order to obtain an SDE, we again need to neglect the third-order derivatives, to obtain
\begin{align}
i\ddt \alpha = \left(\chi\left(|\alpha|^2 -1\right) -\omega_0\right) \alpha \, . \label{TW_Kerr}
\end{align}
As $\hat{H}_\mathrm{Kerr}$ commutes with $\hat{H}_\omega$ (\eq{Homega}), the QFI with respect to $\omega$ will be conserved, so to make a nontrivial calculation we will consider an alternate parameter encoding. After evolving under $\hat{H}_\mathrm{Kerr}$ for duration $t$,  we encode the parameter $v_0$  by evolving for a period $\Delta t$ under the Hamiltonian
\begin{align}
\hat{H}_v &= \hbar v_0 \hat{Y} \, ,
\end{align}
where $\hat{Y} = \frac{i}{\sqrt{2}}(\ahat-\ahatd)$. Figure \ref{fig:kerrFQ} shows the QFI w.r.t. to the parameter $x_0 = v_0 \Delta t$ calculated via the trajectory method, compared to the exact evolution of the quantum state. For large values of $\chi t_1$, the trajectory method and exact method begin to disagree. We attribute this entirely to the truncated Wigner approximation. To demonstrate this, we calculated  $W(\alpha,t)$ directly from \eq{FP_Kerr}, both with and without the third-order derivative terms, and calculated the QFI via \eq{FQpure}. We found that with the inclusion of the 3rd order terms, the results agree with the exact solution to the Schrodinger equation, and in their absence, we find agreement with the trajectory method. 

\begin{figure}[h]
\begin{center}
\includegraphics[width=\columnwidth]{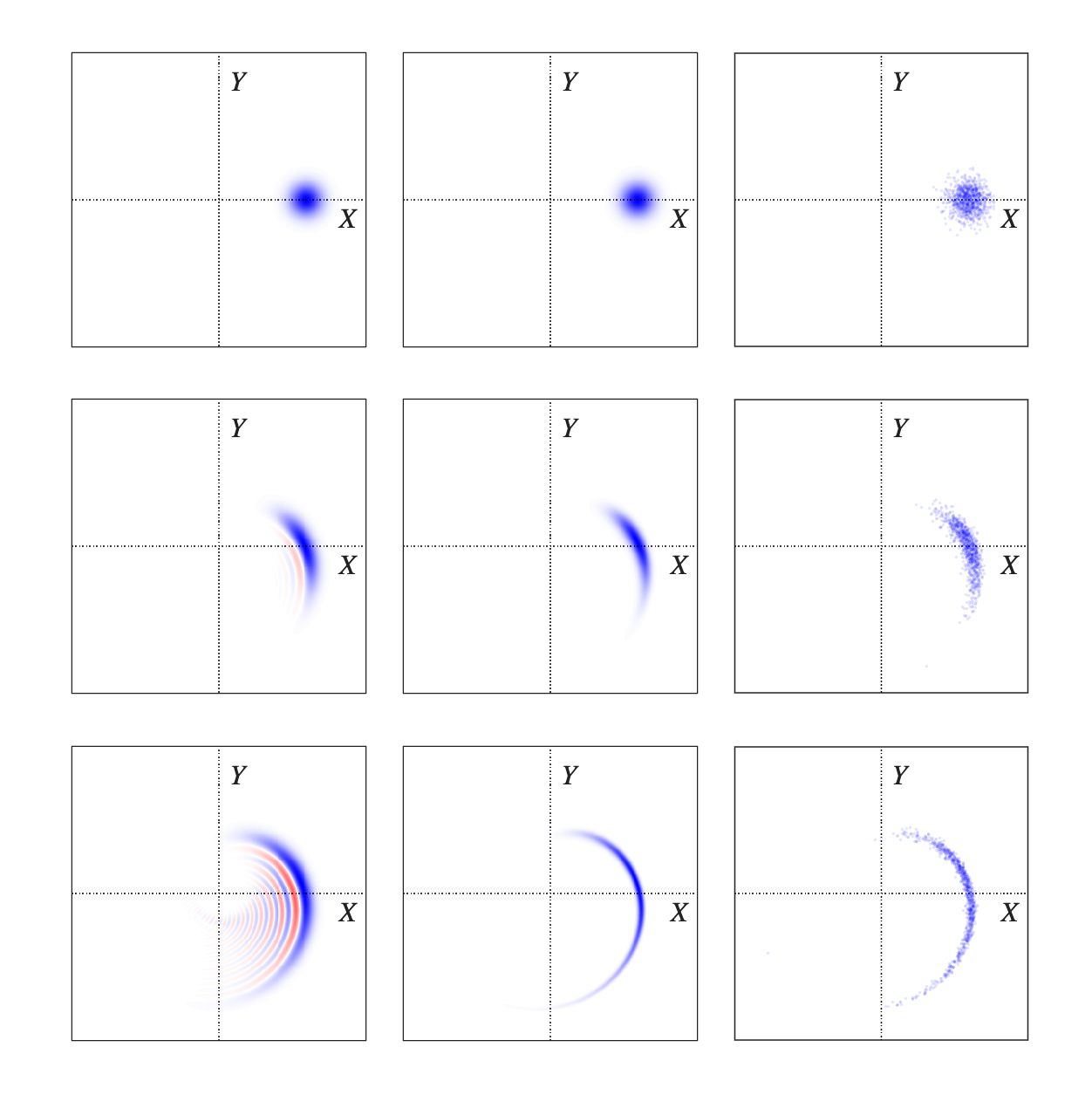}
\caption{$W(\alpha,t)$ with (left column) and without (middle column) the inclusion of the third-order derivative terms, compared to a subset of stochastic trajectories (right column), at $t=0$ (top row), $\chi t=0.03$ (middle row), and $\chi t = 0.07$ (bottom row). The high-frequency fringes in the bottom left frame are the result of negativity (indicated by red) in the Wigner function. An initial coherent state $|\alpha_0\rangle$ with $\alpha_0 = 4$ was used.}
\label{fig:kerrWig}
\end{center}
\end{figure}

To illustrate the effect of the third-order derivative terms in \eq{FP_Kerr}, figure \ref{fig:kerrWig} shows $W(\alpha,t)$ in both cases, as well as a subset of trajectories from the TW method. The appearance of negativity in the full solution corresponds with the time at which the two methods of calculating the QFI begin to disagree.  

The ability to compute the QFI directly from TW simulations adds significant utility to quantum sensing calculations. Without access to the QFI, one would typically rely on a `method-of-moments' (MoM)  estimator to assess the metrological sensitivity. For typical quantum squeezing schemes, a state with increased sensitivity to a displacement along the $X$ axis would be characterized by a decrease in $\mathrm{Var}(\hat{X})$. Specifically, the sensitivity when using this estimator is 
\begin{align}
\delta x_0^2 &= \frac{\mathrm{Var}(\hat{X})}{\left(\partial_{x_0}\langle\hat{X}\rangle\right)^2} = \mathrm{Var}(\hat{X}), 
\end{align}
rather than simply $\delta x_0^2 = 1/F_Q$. Figure \ref{fig:kerrWig} compares $1/\delta x_0^2 = 1/\mathrm{Var}(\hat{X})$ to $F_Q$, and shows that the MoM estimator fails to predict any increase in sensitivity, as $\mathrm{Var}(\hat{X})$ does not decrease due to the state preparation dynamics. As such, the method-of-moments estimator predicts no improvement in sensitivity due to the state preparation. In this example, the QFI, rather than a moment-based sensitivity metric, is essential for accurately characterizing metrological advantage. 

\section{Conclusion}
In this work, we have introduced a trajectory‑based method for estimating the quantum Fisher information (QFI) within the truncated Wigner (TW) framework. Our approach exploits the sensitivity of individual TW trajectories to variations of the metrological parameter, allowing the QFI to be inferred directly from the dispersion of the resulting phase‑space configurations. The numerical results demonstrate that the method performs reliably within the established regime of validity of the TW approximation. As soon as third‑order derivatives—neglected in the truncated evolution—begin to play a significant dynamical role, the method generally overestimates the QFI, providing a clear and physically intuitive diagnostic of the breakdown of both the TW approximation and the QFI reconstruction itself.

It is instructive to compare our approach to the recent method of RouhbakhshNabati \emph{et al}. \cite{RouhbakhshNabati:2025}. Their technique evaluates the QFI for systems undergoing chaotic dynamics by propagating an ensemble of classical trajectories and extracting the parameter sensitivity through derivatives of the classical action. Although not framed explicitly in terms of the TW method, their construction is closely related and could, in principle, be extended to a broader class of systems where TW simulations are applicable. A key distinction, however, lies in the information required by each method: whereas their procedure relies on access to the full classical action accumulated along each trajectory, our method depends solely on the final‑time phase‑space configuration. As a consequence, our approach naturally accommodates situations involving sudden parameter quenches or instantaneous jumps in the dynamics, where the notion of a well‑defined classical action may become ambiguous or inconvenient.

Overall, the method presented here provides a practical and broadly applicable tool for estimating QFI within TW simulations, while also offering a transparent indicator of when the approximation ceases to be reliable.

\section{ACKNOWLEDGMENTS}
The authors would like to acknowledge useful conversations with Zain Mehdi, Jessica Eastman, Stuart Szigeti, Sam Nolan, Nicholas Bohlsen, and James Gardner. This work was supported by funding through an Australian Research Council Future Fellowship, Grant No. FT210100809, ARC Discovery project DP230101685, and Australian government Department of Industry, Science, and Resources via the Australia-India Strategic Research Fund (AIRXIV000025) . This research was undertaken with the assistance of resources and services from the National Computational Infrastructure (NCI), which is supported by the Australian Government. The Australian National University is situated on land traditionally owned by the Ngunnawal people. 

\bibliography{simon_bib_TW.bib}

\end{document}